\documentclass[review]{elsarticle}

\usepackage{lineno,hyperref}

\usepackage{graphicx}
\usepackage{balance}

\usepackage{amsmath}
\usepackage{amsthm}
\usepackage{amsfonts}
\usepackage{algorithm}
\usepackage{algorithmicx}
\usepackage{algpseudocode}
\usepackage{subfigure}
\usepackage{color}
\usepackage{epstopdf}
\usepackage{url}

\usepackage{wrapfig,psfig,epsf}
\usepackage{bbding}

\newtheorem{example}{Example}
\modulolinenumbers[5]

\journal{Journal of \LaTeX\ Templates}

\begin{document}

\begin{frontmatter}

\title{Approximate Query Processing for Group-By Queries based on Conditional Generative Models}


\author[mymainaddress]{Meifan Zhang}

\author[mymainaddress,mysecondaryaddress]{Hongzhi Wang}
\cortext[mycorrespondingauthor]{Corresponding author}
\ead[url]{wangzh@hit.edu.cn}

\address[mymainaddress]{Department of Computer Science and Technology, Harbin Institute of Technology}
\address[mysecondaryaddress]{Peng Cheng Laboratory, Shenzhen, China}

\begin{abstract}
The Group-By query is an important kind of query, which is common and widely used in data warehouses, data analytics, and data visualization. Approximate query processing is an effective way to increase the querying efficiency on big data. The answer to a group-by query involves multiple values, which makes it difficult to provide sufficiently accurate estimations for all the groups. Stratified sampling improves the accuracy compared with the uniform sampling, but the samples chosen for some special queries cannot work for other queries. Online sampling chooses samples for the given query at query time, but it requires a long latency. Thus, it is a challenge to achieve both accuracy and efficiency at the same time. Facing such challenge, in this work, we propose a sample generation framework based on a conditional generative model. The sample generation framework can generate any number of samples for the given query without accessing the data. The proposed framework based on the lightweight model can be combined with stratified sampling and online aggregation to improve the estimation accuracy for group-by queries. The experimental results show that our proposed methods are both efficient and accurate.
\end{abstract}

\begin{keyword}
Approximate query processing\sep Group-By Queries\sep Sampling \sep Conditional Generative Model
\end{keyword}

\end{frontmatter}

%\linenumbers

\section{Introduction}
Querying on big data is costly and suffers from long latency. Approximate query processing (AQP) is proposed to increase the efficiency of querying on big data. Aggregation queries can be estimated by the AQP methods. The group-by aggregation queries plays an important role in interactive data analytics~\cite{DBLP:conf/sigmod/DingHCC016, DBLP:conf/edbt/RoschL09}.%\footnote{(done)any reference?}.

A Group-By query is in the following form:

\texttt{SELECT $A$, $AGGR(B)$ FROM $D$ GROUP BY $A$;}

In this template, $A$ is the Group-By attribute, $B$ is the aggregation attribute, and $AGGR$ is an aggregation function such as the COUNT, SUM, AVG, VAR, and STD.

A group-by query groups the data according to some columns and computes the aggregation of each group. The result of a group-by query includes multiple values. The answer of a group-by aggregation query can be easily transformed into a visualization, such as a bar-chart, a pie-chart, and a histogram, which is useful in data analytics for decision-making~\cite{DBLP:journals/pvldb/KimBPIMR15,DBLP:conf/sigmod/DingHCC016}.

%Data visualization is useful for decision makings, it converts data information into physical visions for observation. Data visualization for big data is a daunting task due to the huge amount of data. Approximate visualization makes a trade-off between the quality of the visualization and efficiency. Approximate visualization is acceptable as long as the approximate result does not affect the decision making based on it. Besides, small errors cannot be recognized by human eyes. In this paper, we focus on visualization based on approximate query processing (AQP).
%AQP approaches are well studied in data community, it aim at estimating the query result with low latency. The aggregation queries with \textit{Group By} can be easily transformed in to visualizations.
Sampling-based AQP~\cite{DBLP:conf/sigmod/DingHCC016,DBLP:conf/sigmod/ChaudhuriDK17,DBLP:journals/vldb/OrrBS20,DBLP:conf/sigmod/HellersteinHW97,DBLP:conf/sigmod/0001WYZ16,DBLP:journals/pvldb/KimBPIMR15,DBLP:conf/icde/NguyenSPXST20} is the most widely used AQP method, but sampling-based AQP for group-by queries is a challenging task. The answer to a group-by query includes multiple values, and it is difficult to get sufficiently accurate estimations for all the groups at the same time~\cite{DBLP:conf/sigmod/DingHCC016, DBLP:conf/icde/NguyenSPXST20}.The sampling-based AQP for group-by queries has problems of either inaccurate, inflexible, or inefficient.
The sampling methods are mainly divided into two categories: off-line sampling and on-line sampling.

The off-line sampling means creating samples before queries come.
Uniform sampling choose each data as a sample with equal probability. It does not work well for group-by queries, since some small groups cannot get sufficient samples from the uniform sampling~\cite{DBLP:conf/icde/NguyenSPXST20, DBLP:journals/dase/LiL18}. Thus, the estimation accuracy of each group varies greatly. Some extremely rare groups will even be missing in the uniform samples.
Stratified sampling~\cite{DBLP:conf/sigmod/AcharyaGP00, DBLP:conf/vldb/GantiLR00, DBLP:journals/tods/ChaudhuriDN07, DBLP:conf/cidr/SidirourgosKB11, DBLP:conf/eurosys/AgarwalMPMMS13, DBLP:conf/icde/NguyenSPXST20} is a way to improve the accuracy. It samples from each group with different probabilities. The stratified sampling usually relies on the workload or some prior knowledge of the distribution. Some methods set the sampling probabilities with the goal of improving the estimation accuracy of the queries in the workload, but the stratified samples chosen for some special queries does not work well for other queries. For example, if the aggregated attribute of a given query is not involved in the workload, the off-line stratified samples chosen based on the workload do not benefit the given query. Preparing samples for all the columns require a huge storage.

The online sampling~\cite{DBLP:conf/sigmod/HellersteinHW97, DBLP:journals/pvldb/KimBPIMR15} means creating samples for the given query after the query comes.
For the queries unrelated to the workload or the ad-hoc queries, the off-line samples has little help. They need online samples chosen for the given queries to improve the accuracy. For low-selectivity queries, choosing enough off-line samples is expensive, since only a few tuples satisfy the predicate of a low-selectivity query. Online-sampling can choose enough samples for the given query predicate to improve the accuracy without maintaining a large off-line sample. Even though online sampling can improve the accuracy for the given query, its drawback is also significant. Sampling from big data at query time means frequently accessing the raw data and a huge I/O cost~\cite{DBLP:conf/sigmod/DingHCC016} .
That is, online-sampling for the given query results in high latency in the query time, which is unacceptable in interactive analysis.

We summarize the pros and cons of these methods.
Uniform sampling is simple, but it cannot provide sufficiently accurate estimation for the rare groups due to the uneven distribution of different groups. Stratified sampling is more accurate, but it cannot flexibly respond to the variation of the query for the reason that the stratified samples are chosen to optimize the accuracy of some given aggregations or the queries in the workload. Online sampling can flexibly get samples for the given query, but it suffers from long latency caused by sampling from big data at query time. Thus, these methods cannot achieve the accuracy, flexibility and efficiency at the same time.

Thus, we attempt to get online stratified samples to improve the estimation accuracy of a group-by query, and avoid the cost of sampling from data at query time. As we mentioned before, the stratified sampling benefits the group-by queries, but the off-line samples cannot cope well with changes in aggregation attributes. Online sampling draws samples for the given query at the cost of the high latency. If we can get stratified samples for the given query without accessing the data, both efficiency and accuracy can be achieved.

However, getting stratified samples without accessing the data is still a challenging task. The generative model such as the GAN can learn the data distribution and generate samples for the entire dataset. However, it cannot generate targeted stratified samples to improve the estimation accuracy of group-by queries. Thus, it still suffers from the sampling error like the uniform sampling.
We thought of using the conditional generative models instead of the raw data to generate stratified samples. A conditional generative model, such as the conditional generative adversarial network (CGAN)~\cite{DBLP:journals/corr/MirzaO14, DBLP:journals/ejwcn/QinJ18, DBLP:conf/embc/Luo18}, can generate images of a given type.
We want to enable the conditional generative model to generate samples for each given group. Thus, the rare groups can easily get enough samples to improve the accuracy. We convert the group-by attribute values to be the labels, and adopt the conditional generative model to learn the conditional data distribution. After the model is well tuned, it allows targeted sample generation for a given group. Thus, we can directly get the stratified samples from the model without accessing the raw data.

The proposed method can achieve both high efficiency and accuracy.
On the one hand, the samples can be obtained without scanning the data, which greatly reduces the response latency.
On the other hand, the model can be used to generate any number of samples for the given group, thus, small groups can easily get enough samples to improve the accuracy.

The proposed method can be easily combined with many kinds of stratified sampling methods to allocate the sample size for each group. The allocation can be computed according to some statistics, such as the frequency, mean, and the variation of each group. These statistics can be computed in a single pass through the data in advance.

Based on the proposed sample generation method, it is easy to generate enough samples for each group. However, for the group-by query with a predicate, the generated samples still need to be filtered by the predicate. In order to obtain enough samples that satisfy the predicate, a large number of useless samples need to be generated, which will cause a waste of time. Thus, approximate query processing for a group-by query with a predicate is also a challenging task.
%\footnote{(done)need connection}
%We also consider the Group-By queries with a predicate.

The group-by query with a predicate is in the following form:

\texttt{SELECT $A$, $AGGR(B)$ FROM $D$ WHERE Condition GROUP BY $A$;}

The predicate is a filtering condition, such as ``$l\le C\le r$'', and the answer to this kind of query is the aggregation of each group after filtering through the condition. If the tuples in the data rarely satisfy the predicate, in other words, it is a low-selectivity query, then there are also very few tuples that satisfy the predicate in the generated samples. We can simply generate more samples to get enough ones satisfying the predicate, but it is not an ideal method.

%\footnote{(done)connect to above paragraph and discuss the motivations}
%\textcolor[rgb]{0.00,0.07,1.00}{
In order to reduce the waste caused by generating useless samples, we attempt to enable the generative model to generate targeted samples for the given predicate. However, it is not trivial to label the data with the predicate, since a data belongs to all the ranges that cover it. In order to label the tuples with range information, we partition the data with a histogram. Thus, each tuple belongs to one bucket. We train the conditional generative model with both the bucket labels and group labels. Once the model is well-tuned, it can generate samples for the given bucket and group. This method improves the sample quality for the low-selectivity queries.

We make the following contributions in this paper.

(1) We proposed a sample generation framework for group-by queries based on the conditional generative model. The proposed framework can generate any number of samples for the given group. It can be combined with stratified sampling and online aggregation to improve the estimation accuracy.

(2) We proposed a sample generation framework for the group-by queries with predicates. The proposed framework can generate any number of samples for a sub-range of the data, which benefits the estimations of the low-selectivity queries.

(3) The experimental results show that our method can efficiently generate stratified samples to improve the accuracy, iteratively narrow down the estimation error like the online aggregation, and work well for the low-selectivity queries. We also compare the proposed method with the state-of-the-art machine-learning-based method DBEst. The experimental results show that our method is more accurate while occupying less space.

The remaining parts of this paper are organized as follows. In Section~\ref{section:relatedwork}, we survey the related works for this paper. In Section~\ref{section:GroupBy}, we introduce the sample generation framework for the group-by queries and the methods of combining a stratified sampling and online aggregation with the proposed method to improve the accuracy. In Section~\ref{section:GroupByWithPredicates}, we introduce the method to generate stratified samples for the group-by queries with predicates. In Section~\ref{section:Experiments}, we show the experimental results. Finally, we conclude this study in Section~\ref{section:conclusion}.

\section{Related Work}\label{section:relatedwork}

%Visualization is useful in data exploration to find the insight and statistic of data for data analysis or decision-makings. Visualization on big data is a challenging task. Instead of generating the exact visualizations of the whole data set, we prefer getting the approximate visualizations based on queries.
%In this paper, we focus on the approximate visualization based on AQP for the aggregation queries with \textit{Group By}, since they are the most important class of the ad-hoc queries~\cite{DBLP:conf/sigmod/DingHCC016}. The results of the \textit{Group By} queries can be easily transformed into visualizations such as the bar-charts, pie-charts, and histograms.

%sampling based AQP with GROUP BY
Sampling-based approximate query processing method~\cite{DBLP:conf/sigmod/DingHCC016, DBLP:conf/sigmod/ChaudhuriDK17, DBLP:journals/vldb/OrrBS20, DBLP:conf/sigmod/HellersteinHW97, DBLP:conf/sigmod/0001WYZ16} is the most widely used AQP approach, since it is simple and can be adaptive to most general queries. Many of works have been proposed in the past.
The uniform sampling does not work well for the high-skew data. The answers of group-by queries includes the aggregation in each group. The distribution of the groups is not always uniform. The uniform sampling cannot collect enough samples for the rare groups. It demands a big sample to get a satisfactory estimation for a low-selectivity query or the high-skew data. Stratified sampling is designed to improve the accuracy, but it requires some foreknowledge such as the expected query workload and the distribution of the aggregated attribute. In addition, the stratified sample only benefits the given specific query~\cite{DBLP:journals/vldb/OrrBS20}. Estimating queries over rare sub-populations based on the SAQP is still challenging~\cite{DBLP:conf/sigmod/ChaudhuriDK17}. Sample+Seek~\cite{DBLP:conf/sigmod/DingHCC016} proposed a new metric called `distribution precision' to measure the accuracy of the visualization. It developed a measure-biased sampling method, which picks a row with probability proportional to its value on the measure attribute.
CVOPT~\cite{DBLP:conf/icde/NguyenSPXST20} is a recent stratified sampling method which allocate the sample size for each group according to the coefficients of variation, and the experimental results show that it achieves lower maximum relative error compared with uniform sampling, congressional sampling~\cite{DBLP:conf/sigmod/AcharyaGP00} and the stratified sampling method proposed in reference~\cite{DBLP:conf/edbt/RoschL09}.
Online aggregation~\cite{DBLP:conf/sigmod/HellersteinHW97, DBLP:conf/sigmod/0001WYZ16} is another way for AQP. It iteratively accesses more samples to improve the estimation accuracy. The online aggregation draws samples at query time, thus, it suffers from the high latency caused by sampling from big data.

%progressive
Progressive methods provide a way to iteratively improve the quality of estimations.
IFOCUS~\cite{DBLP:journals/pvldb/KimBPIMR15} and INCVISAGE~\cite{DBLP:journals/pvldb/RahmanAKBKPR17} incrementally improve the approximate estimations. IFOCUS~\cite{DBLP:journals/pvldb/KimBPIMR15} aims at guaranteeing the order of estimations of different groups. It keeps increasing the sample size until the confidence interval of each group does not overlapping with others. INCVISAGE~\cite{DBLP:journals/pvldb/RahmanAKBKPR17} is designed to avoid the incorrect intermediate visualizations generated by the incremental sampling-based method such as the SampleAction~\cite{DBLP:conf/chi/FisherPDs12}. The INCVISAGE is conservative at the beginning, and it keeps splitting one segment into two in order to refine the visualization. Different from the SampleAction~\cite{DBLP:conf/chi/FisherPDs12}, it finally reveals the features of the eventual visualization.

%synopsis based
Synopses such as histograms, sketches, wavelets are also used for AQP in the past, but most of them only store some summarizations such as count, density, average, for the sake of lightweight. The synopses-based AQP methods have been summarized in the surveys ~\cite{DBLP:journals/ftdb/CormodeGHJ12, DBLP:journals/dase/LiL18}.
A group-by query has a multi-dimensional result rather a simple value. Therefore, adopting the synopses to estimate the group-by queries will cost a huge storage.

Making use of previous answers to improve the estimations of the new queries is another typical idea. An AQP formulation in reference~\cite{DBLP:journals/pvldb/GalakatosCZBK17} answers new queries using the old results according to the probability theory. It also builds the tail index for the rare subpopulation.

%machine learning
Machine learning methods have been adopted in query processing in recent years~\cite{DBLP:conf/sigmod/ParkTCM17, DBLP:conf/sigmod/MaT19, DBLP:conf/icde/Thirumuruganathan20, DBLP:journals/pvldb/HilprechtSKMKB20}. \textit{Database Learning}~\cite{DBLP:conf/sigmod/ParkTCM17} learns from past answers to improve the new query answers. DeepDB~\cite{DBLP:journals/pvldb/HilprechtSKMKB20} adopted the sum-product network to model the data probability distribution.
EntropyDB~\cite{DBLP:journals/vldb/OrrBS20} uses the principle of maximum entropy to generate a probabilistic representation of the data that can be used to give approximate query answers.
Reference~\cite{DBLP:conf/icde/Thirumuruganathan20} proposed a AQP method using deep generative models to learn data distribution and approximately answer queries by generating samples from the learned models. It reduces the latency, but it still suffers from the sampling error. DBEst~\cite{DBLP:conf/sigmod/MaT19} is a machine-learning-based AQP method, which builds the probability density model and regression model according to samples from data. It uses the model to answer the queries directly. For the group-by queries, it builds a model for each group, which will increase the cost of storage.

\section{AQP for Group-By Queries based on Conditional Generative Model}\label{section:GroupBy}
In this section,  we introduce the framework of sample generation for group-by queries based on a conditional generative model.
%\footnote{(done)give the roadmap}\textcolor[rgb]{0.00,0.07,1.00}{
The sample generation framework is introduced in section~\ref{section:CWGAN_framework}.
We label each tuple with its group-by attribute value, and feed the data with labels to the conditional generative model. After training, the generator of the model can generate samples conditional on the given group that are similar to real samples of that group.
In section~\ref{section:CWGAN_AQP}, we introduce methods to generate uniform samples and stratified samples based on the sample generation framework. We also provide methods for online aggregation based on the proposed sample generation framework. 
%In section~\ref{section:CWGAN_framework}, we introduce the framework of sample generation for group-by queries based on a conditional generative model. In section~\ref{section:CWGAN_AQP}, we introduce methods to combine the conditional generative model with uniform sampling, stratified sampling, and online aggregation.

\subsection{Framework}\label{section:CWGAN_framework}
In this part, we introduce the conditional sample generation framework for group-by queries.
We attempt to generate the stratified samples to improve the accuracy. There are many stratified sampling methods with different optimization goals, and they have different allocations for the sample size of each group. We do not want to limit the sample generation framework to one kind of stratified sampling methods, but to flexibly generate stratified samples according to different accuracy optimization goals, which can increase the applicability of the sample generation framework. According to our requirements, we hope that the sample generation model can generate any number of samples for any given group. %}
%\footnote{(done)discuss the motivation from the requirements and derive the framework}

\begin{figure}
\caption{The framework of sample generation for group-by queries based on CWGAN}
\centering
\includegraphics[width = 0.6\textwidth]{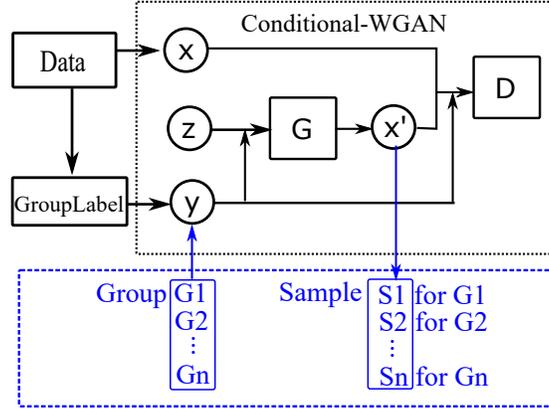}
\label{Fig:CWGAN_GenSampleFramework}
\end{figure}

Fig~\ref{Fig:CWGAN_GenSampleFramework} introduces the framework of generating samples for group-by queries based on a conditional generative model CWGAN.
CWGAN is the combination of the Conditional Generative Adversarial Nets~\cite{DBLP:journals/corr/MirzaO14} (CGAN) and Wasserstein Generative Adversarial Nets~\cite{DBLP:journals/corr/ArjovskyCB17} (WGAN). CGAN is an extension of the Generative Adversarial Nets (GAN), which adds conditional settings to both the generator and the discriminator of the GAN, and thus can generate data for the given condition. WGAN is a variation of GAN, which can improve the stability of learning, and get rid of problems of GAN like mode collapse.

We adopt the CWGAN to generate samples for group-by queries in this framework.

The first step of this framework is to label the data. Since we want enable the model to generate samples for the given group, we label the data with the group attribute values. We index the values and encode their indices by the one-hot encoding, which is a type of vector representation widely used in many branches of computer science, especially machine learning and digital circuit design. The encoded group-by attribute values will be regarded as the labels for the training data. We will an example to illustrate the encoding later.

The next step of this framework is to train the CWGAN with the labeled data. It feeds each data $x$ and its corresponding label $y$ to the model, and trains both the generator $G$ and the discriminator $D$ with the labels. After the model is well-tuned, the generator can generate a sample $x'$ for a given label similar to the real sample $x$. That is, the model can generate any number of samples for each targeted group.

%Before training the model, we first need to encode the values of the group-by attribute. We index the values and encode their indices by the one-hot encoding, which is a type of vector representation widely used in many branches of computer science, especially machine learning and digital circuit design. We use the following example to show that how to encode the group-by attribute values.
%\footnote{above description is fine, but suggest to description in the same order of processing}
We use the following example to show that how to encode the group-by attribute values.
\begin{example}
Table~\ref{table:CWGAN_encoding} shows the way to encode the values of the group-by attribute `gender'. The attribute has two distinct values, `male' and `female'. We index the values with `0' and `1', respectively. Since there are only two distinct values, the one-hot encoding adopts a two-digit register. The `male' and `female' are finally encoded as `1 0' and `0 1'.
\end{example}

\begin{table}[htbp]
\caption{Example of encoding the conditional label}\label{table:CWGAN_encoding}
\centering
\begin{tabular}{ccc}
\hline
gender & index & One-hot encoding \\
\hline
male & 0 & 1 0 \\
\hline
female & 1 & 0 1 \\
\hline
\end{tabular}
\end{table}

After training the conditional generative model, we can use it to generate any number of samples for each group without accessing the data, suggesting that sampling can be conducted independent of data. We can flexibly allocate the sample size of each group to improve the estimation accuracy.

We use the following example to show that how to use the model to generate samples for the given group.

\begin{example}
Suppose we require five samples for the group ``male''. The values of attribute ``gender'' are encoded as shown in Table~\ref{table:CWGAN_encoding}, and ``male'' is encoded as `1 0'. Suppose the latent layer has two dimensions, and the noise $z$ is generated from the normal distribution $z\sim{N(0,1)}$. We need to feed five two-dimensional noises and the group label shown in Table~\ref{table:CWGAN_generator_input} to the generator, and the output of the generator is the five samples for group ``male''.

\begin{table}[htbp]
\caption{Input of Generator}\label{table:CWGAN_generator_input}
\centering
\begin{tabular}{cc}
\hline
noise $z$ & group label\\
\hline
0.0223 -0.7296  & 1 0 \\
\hline
1.3867 0.8051  & 1 0 \\
\hline
-0.7237 -0.6542  & 1 0 \\
\hline
0.7410  0.5340  & 1 0 \\
\hline
1.2362  0.9117 & 1 0 \\
\hline
\end{tabular}
\end{table}

\end{example}

\subsection{CWGAN-based Approximate Group-by Queries}\label{section:CWGAN_AQP}

Since the CWGAN-based sampling can generate samples for each group without accessing the data, it can be easily combined with various sampling methods. In this section, we introduce how to combine CWGAN with the uniform sampling method, a stratified sampling method CVOPT, and the online aggregation. Generating uniform samples can restore the data distribution, and generating stratified samples can improve the estimation accuracy for group-by queries. Generating samples for online aggregation can iteratively increase the accuracy and allow the user to stop the processing when the estimation is accurate enough.

\subsubsection{CWGAN-based uniform sampling}
The proposed sample generation approach based on CWGAN can generate uniform samples by allocating the sample size of each group according to the frequency of each group.

The sampling probability of each data is the same in the uniform sampling. It means that the number of samples for each group respects to the frequency of this group in the whole dataset. The algorithm~\ref{algorithm:CWGAN_UniformSample} introduces how to generate uniform samples for the data. The frequency of each group [$COUNT_1$, $COUNT_2$,... , $COUNT_{g}$] is known in advance, and the algorithm allocates the $SampleSize$ according to these frequencies.
 The $Random(z, SampleSize)$ means to get $SampleSize$ random noises $z$ (Line 3), and the $Repeat(\operatorname{One-hot(i)}, SampleSize)$ means to repeat the label of the $i$th group $SampleSize$ times (Line 4). It then feeds both the noises and the labels to the generator, and get the samples $S_i$ for the $i$th group (Line 5).

\begin{algorithm}
\caption{CWGAN-based uniform sampling}\label{algorithm:CWGAN_UniformSample}
\textbf{Input: Generator $G$, TotalSampleSize $m$, Number of Groups $g$, Frequencies of each group [$COUNT_1$, $COUNT_2$,..., $COUNT_{g}$]}\\
\textbf{Output: Samples for each group $S_1$, $S_2$,..., $S_g$ }
\begin{algorithmic}[1]
\For {$i$ in 1 to $g$}
    \State $SampleSize = m\cdot\frac{COUNT_i}{\sum_{j=1}^{g}COUNT_j}$
    \State $noise\leftarrow Random(z, SampleSize)$
    \State $label\leftarrow Repeat(\operatorname{One-hot(i)}, SampleSize)$
    \State $S_i\leftarrow G(noise, label)$
\EndFor
\State \textbf{Return:} $S_1$, $S_2$,..., $S_g$
\end{algorithmic}
\end{algorithm}

Under the premise that the frequency of each group is known in advance, the generative model can generate uniform samples to restore the distribution of the data.

\subsubsection{CWGAN-based stratified sampling}
The proposed sample generation framework based on CWGAN can generate stratified samples by allocating the sample size of each group according to the standard deviation and mean of each group.

In this part, we combine the CWGAN with a stratified sampling method, CVOPT, which allocates the samples according to the coefficient of variation of each group. The coefficient of variation (CV) is defined as $CV[X]=\frac{standard\_deviation[X]}{mean[X]}$. The CVOPT is proved to construct a stratified sample optimizing the $l_2$ norm of the CVs of different answers. Algorithm~\ref{algorithm:CWGAN_CVOPTSample}~introduces the way of combining CWGAN with CVOPT to generate stratified samples. The difference of this algorithm from Algorithm~\ref{algorithm:CWGAN_UniformSample} is the $SampleSize$ allocation for each group (Line 2).

%以AVG分组查询为例，第i个组的估计值为：
%\begin{align}
%  EST_i= AVG_i\pm\lambda\frac{\sigma_i}{\sqrt{|S_i|}},
%\end{align}

%其中$AVG_i$表示第$i$个组内聚集属性上的均值，$VAR_i$是第$i$个组内聚集属性上的方差，$S_i$为第$i$个组的样本量，$\lambda$是和置信区间关的函数，例如，$\lambda=1.96$时上述近似结果置信度为95\%。
%CVOPT通过使各组上的$CV$的
\begin{algorithm}
\caption{CWGAN-based stratified sampling}\label{algorithm:CWGAN_CVOPTSample}
\textbf{Input: Generator $G$, TotalSampleSize $m$, Number of Groups $g$, Means of each group [$\mu_1$, $\mu_2$,..., $\mu_{g}$], Standard deviation of each group[$\sigma_1$, $\sigma_2$,..., $\sigma_{g}$]}\\
\textbf{Output: Samples for each group}
\begin{algorithmic}[1]
\For {$i$ in 1 to $g$}
    \State $SampleSize = m\cdot\frac{\sigma_i/\mu_i}{\sum_{j=1}^{g}\sigma_j/\mu_j}$
    \State $noise\leftarrow Random(z, SampleSize)$
    \State $label\leftarrow Repeat(\operatorname{One-hot(i)}, SampleSize)$
    \State $S_i\leftarrow G(noise, label)$
\EndFor
\State \textbf{Return:} $S_1$, $S_2$,..., $S_g$
\end{algorithmic}
\end{algorithm}

Since the CVOPT is reported to reduce the maximum relative error compared with the uniform sampling. The samples generated according to the allocation of CVOPT will also increase the accuracy, on the assumption that the generative model is reliable.
This algorithm requires the statistics including the standard deviation and the mean of each group, and these statistics can be easily computed with one-pass of data in advance.

\subsubsection{CWGAN-based Online-Aggregation}
Since the generative model can generate any number of samples for a given group, we can apply the generated samples for online-aggregation to iteratively improve the accuracy in an interactive way.
Online-aggregation keeps improving the accuracy and computing the running confidence intervals by increasing the sample size. It allows the user to stop the processing of each group when the estimation is accurate enough.

Algorithm~\ref{algorithm:CWGAN_OnlineAggregation} introduces the way of combining CWGAN with the online-aggregation. It adopts the width of confidence interval as the termination of the algorithm. Once the confidence interval of a group is narrower than the object width, the algorithm will stop generating samples for that group. In this algorithm, the $ActiveGroup$ is used to store the indices of the groups requiring more samples.

This algorithm adopts the Round-Robin to iteratively generate one sample for each group, and it is convenient to observe the accuracy of each group after generating the same number of samples.
Since the model is easy to generate samples for the given group, we can use it to generate samples for one group until its termination is reached. %It can also be used to generate samples for each group in parallel.

\begin{algorithm}
\caption{CWGAN-based Online-Aggregation}\label{algorithm:CWGAN_OnlineAggregation}
\textbf{Input: Generator $G$, Number of Groups $g$, Object Confidence Interval[$I_1$, $I_2$,..., $I_{g}$]}\\
\textbf{Output: Samples for each group $S_1$, $S_2$,..., $S_g$}
\begin{algorithmic}[1]
\State $ActiveGroup\leftarrow [1,2,...,g]$
\State Initialize the $S_1$, $S_2$,..., $S_n$ to be empty sets.
\While {$ActiveGroup\neq\emptyset$}
    \For {$i$ in $ActiveGroup$}
        \State $noise\leftarrow Random(z, 1)$
        \State $label\leftarrow Repeat(\operatorname{One-hot(i)}, 1)$
        \State $s\leftarrow G(noise, label)$
        \State $S_i=S_i\cup s$
        \State $CurrentI_i\leftarrow$ Confidence Interval of $S_i$
        \If {$CurrentI_i<I_i$}
            \State $ActiveGroup\leftarrow ActiveGroup\setminus i$
        \EndIf
    \EndFor
\EndWhile
\State \textbf{Return:} $S_1$, $S_2$,..., $S_g$
\end{algorithmic}
\end{algorithm}

Our conditional sample generation framework can iteratively improve the accuracy by generating more samples like the online aggregation. But the sample generation does not require random I/O accesses, and thus largely reduces the latency at query time.

The proposed framework can also be combined with the other online sampling method such as the IFOCUS, which keeps increasing the sample size of the groups whose confidence interval overlap with other groups, until the confidence interval of each group does not overlap with others. The algorithm of CWGAN-based IFOCUS can be obtained by modifying the termination of each group in Algorithm~\ref{algorithm:CWGAN_OnlineAggregation} to the non-overlap of the confidence intervals.

%\footnote{\textcolor[rgb]{1.00,0.00,0.00}{this section is clear. suggest to add the example and theoretical analysis}}

\section{CWGAN-based Sampling for Group-By Queries with predicates}\label{section:GroupByWithPredicates}

In this section, we propose method to generating the samples for the group-by queries with predicates.

The group-by query with a predicate is in the following form:

\texttt{SELECT $A$, $AGGR(B)$ FROM $D$ WHERE Condition GROUP BY $A$;}

The ``Condition'' in the predicates is in the form like ``$l\le C\le r$''. This kind of queries can be estimated in two ways. The first is to estimate a query based on the samples chosen from the pre-computed samples according to the predicate. But the estimations based on the chosen samples are not satisfactory for the low-selectivity queries, since only a small part of the samples satisfying the predicate contribute to the estimations. Another way is to separately choose samples for each query. The samples can be chosen from the data filtered by the query predicate. This method will undoubtedly improve accuracy, but it requires sampling from the entire dataset several time. Query time sampling will cause unacceptable long latency.

Even though the conditional sample generation framework can generate sample without accessing the data, we do not want to waste time of generating sample without contribution to the estimation. We attempt to enable the conditional generative model to generate samples for the given query predicate. However, the conditional generative model cannot be simply adopted to solve this problem. The reason is that, we cannot directly set a range label for each tuple like setting the group labels, since each tuple belongs to all the ranges covering it.

In order to give each data a clear range label, we consider bucketing the data range through a histogram. A histogram can divide the data into several disjoint buckets. Each bucket contains a sub-range of data, and each tuple only corresponds to one bucket. We add a bucket label for each tuple according to the index of its located bucket. Thus, we can train the model by feeding the data with both group labels and bucket labels to the conditional generative model. Finally, the generator can generate samples for a given bucket and a given group.

\begin{figure}
\centering
\includegraphics[width = 0.6\textwidth]{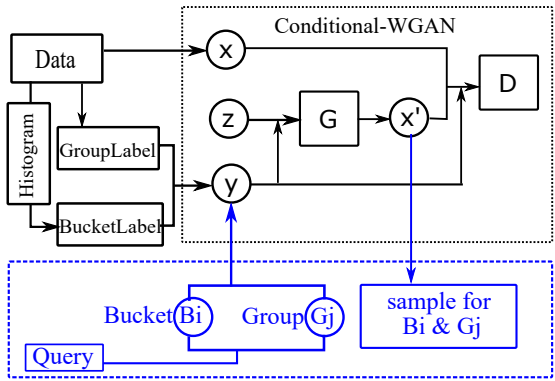}
\caption{The framework of sample generation for group-by queries with predicates}
\label{Fig:CWGAN_ConditionalGenSampleFramework}
\end{figure}

The framework of sample generation for the group-by queries with predicates is shown in Fig~\ref{Fig:CWGAN_ConditionalGenSampleFramework}. It differs from the framework proposed in the previous section in that it uses two kinds of labels to train the conditional generative model. We also adopt the CWGAN as the conditional generative model. The input of the generator includes the random noises, bucket labels, and group labels.

In this framework, we partition the data with a histogram. We do not limit which kind of histogram to use.
Different histogram schemes, such as the equi-width histogram, equi-depth histogram, and V-optimal histogram, can be applied to this framework. We prefer the equi-depth histogram, which has the same number of tuples in each bucket. It benefits for both the model training and the sample quality. On the one-hand, it balances the quantity of training data corresponding to each bucket. On the other hand, it limits the number of the generated samples that do not satisfy the query predicate.

Algorithm~\ref{algorithm:CWGAN_SampleforPredicates}~introduces how to generate samples for a given query. In this algorithm, the model generates an equal number of samples for each bucket that intersects the query predicate. It adopts the CVOPT sampling method to allocate the sample size for each group. The mean and standard deviation of the aggregated attribute in each group should be aware in advance.

The algorithm first finds the buckets intersected with the query predicate (Line 1-6). It allocates an equal number of samples for each bucket (Line 8). The $SampleSize$ for each group is allocated according to CVOPT (Line 10). The algorithm then obtains the random noise (Line 11), the $BucketLabel$ (Line 12), and the $GroupLabel$ (Line 13). The generator $G$ finally generates the samples $S_{i,j}$ for the $i$th bucket and $j$th group.

\begin{algorithm}
\caption{Generate samples for group-by queries with predicates}\label{algorithm:CWGAN_SampleforPredicates}
\textbf{Input: Generator $G$, Groups $g$, total sample size$m$, Query $Q$, Buckets $B_1,B_2,...,B_{n}$, $\mu_{i,j}$ and$\sigma_{i,j}$ for each bucket $B_i$ and group $j$}\\
\textbf{Output: Samples $S_{i,j}$}
\begin{algorithmic}[1]
\State $RB\leftarrow\emptyset$
\For {$i$ in 1 to $n$}
    \If {$B_i$ intersect with $Q$}
        \State $RB\leftarrow \cup i$
    \EndIf
\EndFor
\For {$i$ in $RB$}
    \State $BSampleSize\leftarrow m/|RB|$
    \For {$j$ in 1 to $g$}
        \State $SampleSize = BSampleSize\cdot\frac{\mu_{i,j}/\sigma_{i,j}}{\sum_{k=0}^{g-1}\mu_{i,k}/\sigma_{i,k}}$
        \State $noise\leftarrow Random(z, SampleSize)$
        \State $BucketLabel\leftarrow Repeat(\operatorname{One-hot(i)}, SampleSize)$
        \State $GroupLabel\leftarrow Repeat(\operatorname{One-hot(j)}, SampleSize)$
        \State $S_{i,j}\leftarrow G(noise, BucketLabel,GroupLabel)$
    \EndFor
\EndFor
\State \textbf{Return:} Samples $S_{i,j}$, $i\in RB$, $j=1,2,...,g$
\end{algorithmic}
\end{algorithm}

Algorithm~\ref{algorithm:CWGAN_SampleforPredicates}~ can generate targeted samples for the buckets intersecting the query predicate. Though the generated samples still require to be filtered by the predicate, this method greatly reduces the proportion of samples that do not contribute to the estimation accuracy.

%\footnote{\textcolor[rgb]{1.00,0.00,0.00}{again, examples and analysis}}

\section{Experimental Results}\label{section:Experiments}

\subsection{Settings}

\subsubsection{Hardware and Library}

All the experiments were conducted on a laptop with an Intel Core i5 CPU with 2.60GHz clock frequency and 8GB of RAM.
We use Keras, a python deep learning library running on top of TensorFlow, to build the conditional generative models.

\subsubsection{Datasets}
%\footnote{\textcolor[rgb]{1.00,0.00,0.00}{ discuss why to choose these two data sets}}
\noindent \underline{Real dataset}: ROAD dataset \footnote{\url{http://archive.ics.uci.edu/ml/datasets/3D+Road+Network+(North+Jutland,+Denmark)}} was constructed by adding elevation information to a 2D road network in North Jutland, Denmark. The dataset contains 434,874 tuples. This dataset has three attributes `Longitude', `Latitude', and `Elevation'. We added a group-by attribute `GroupID' to this dataset, and partitioned the value range of the `Latitude' into 10 equal sub-range. The value for the added group-by attribute of each tuple is the index of the sub-range which the tuple belongs to. Thus, the modified dataset has four attributes.

\noindent \underline{Synthetic dataset}: GAUSS dataset has 100M tuples and 4 attributes named `x',`y',`z', and `GroupID'. The first two attributes values are composed of 4 two-dimensional gaussian distributions. The mean and covariance of each distribution are as follows:
GAUSS1(mean=[2, 2], cov=[[30, 0], [0, 30]]), GAUSS2(mean=[80, 10], cov= [[80, 0], [0, 30]]), GAUSS3(mean=[10, 40], cov=[[10, 0], [0, 20]]), GAUSS4(mean=[70, 60], cov= [[100, 0], [0, 100]]). The `GroupID' of each tuple is the index of the gaussian distribution to which the tuple belongs. That is, the domain of `GroupID' is \{1, 2, 3, 4\}. The attribute `z' will be adopted as an aggregation attribute in the experiments. The values of `z' of the tuples in the four distributions are chosen uniformly from the ranges [0, 10000], [0, 1000], [0, 100], [0, 10], respectively. The number of tuples in these four distributions are 10M, 20M, 30M, and 40M. That is, the variance of the smaller group is higher.

\subsubsection{Accuracy Metrics}

We adopt two metrics to measure the accuracy of the query estimation. In the following equations, the $g$, $est_i$, and $true_i$ mean the number of groups, the estimated aggregation of the $i$th group, and the true aggregation of the $i$th group, respectively.

Average Relative Error :
\begin{equation}
  ARE = \frac{1}{|\{true_i|true_i>0, i\in [0, g]\}|}\sum_{i\in [0, g], true_i>0}\frac{|est_i-true_i|}{true_i}
\end{equation}

Mean Squared Error :
\begin{equation}
MSE = \frac{1}{g}\sum_{i\in [0, g]}(est_i-true_i)^2
\end{equation}

\subsubsection{Implementations}\label{section:CWGAN-implement}

We adopt the CWGAN as the conditional generative model for our experiments. CWGAN is the combination of the WGAN and CGAN.

The generator of CWGAN has three fully connected hidden-layers, and the number of neurons in each layer is 128, 64, 32.
The discriminator of CWGAN has three fully connected hidden-layers, and the number of neurons in each layer is 128, 64, 32.

For the group-by queries without predicates, the inputs of generator are the random noises and the group labels.
For the group-by queries with predicates, the inputs of the generator include the random noises, the group labels, and the bucket labels.

We set the activation function as RELU~\cite{DBLP:conf/icml/NairH10}, which is a widely used activation function and yields better results compared to Sigmoid and Tanh.
We set the optimizer as RMSprop~\cite{hinton2012neural} according to the recommendation of WGAN. We set the parameters of RMSprop as ($lr$=0.0001, $rho$=0.95). The clip for WGAN is set as [-0.1, 0.1]. The settings of other parameters follow the recommendation in WGAN~\cite{DBLP:journals/corr/ArjovskyCB17}.%\footnote{(done)reference?}.

We just adopt a simple model with a few layers, and the settings mostly follow the WGAN. More layers and more careful tuning may result in a better model. But it will also increase the difficulty of training and the size of the model. Actually, the structure and settings of the model should be modified according to the dataset. In this experiment, we are not committed to find the optimal model, but to test the idea of adopting conditional generative model to approximate group-by queries. The following experimental results verified that the simple model is able to generate good enough samples.

\subsection{Sample Distribution}
In this experiment, we visualize the distribution of the random samples and generated samples.
This experiment is conducted on the ROAD and GAUSS datasets.
We choose 1,000 random samples from each dataset, and generate 1,000 samples based on the generative model for each dataset. In order to restore the original data distribution, the number of samples generated in each group is allocated according to the number of tuples in that group of the whole dataset. That is, we adopt the Algorithm~\ref{algorithm:CWGAN_UniformSample} to generate uniform samples for each dataset.

\begin{figure*}[!h]
	\centering
	\subfigure[Random Sample]{
        \includegraphics[width=0.4\textwidth]{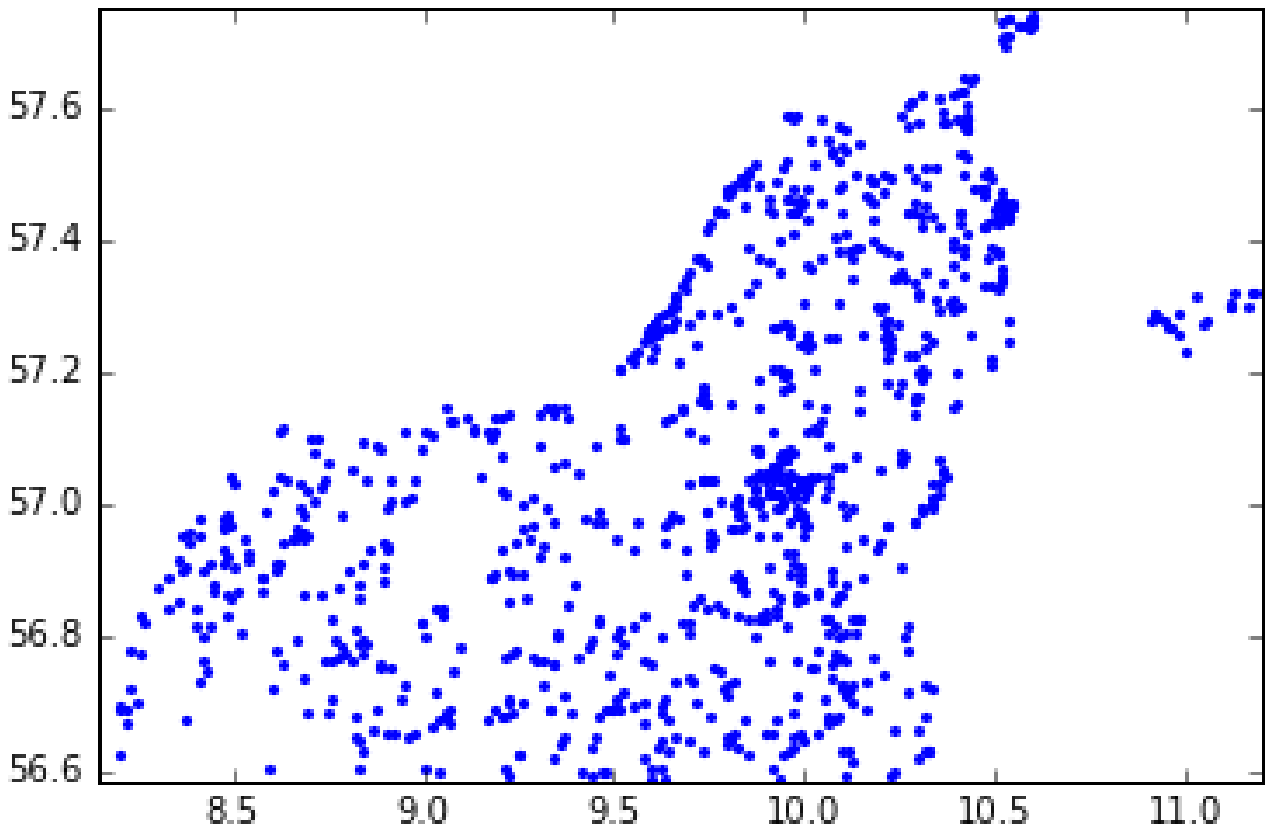}
        \label{fig:CWGAN_RAND_SAMPLE}
	}
    \subfigure[Generated Sample]{
        \includegraphics[width=0.4\textwidth]{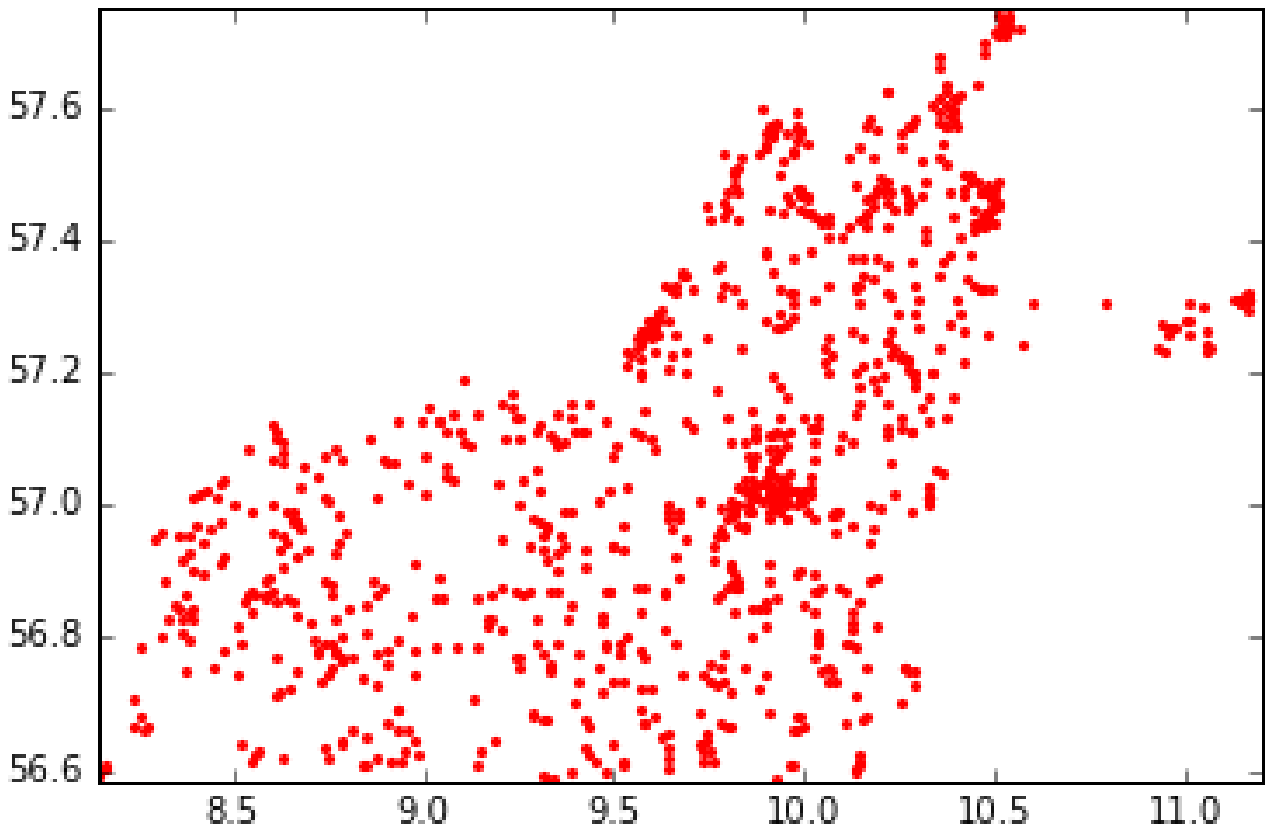}
        \label{fig:CWGAN_GEN_SAMPLE}
	}
	\caption{Random Sample \&.Generated Sample (ROAD dataset)}
	\label{fig:RandomSample&GeneratedSample}
%\vspace{-0.3cm}
\end{figure*}

\begin{figure*}
	\centering
	\subfigure[Random Sample]{
        \includegraphics[width=0.4\textwidth]{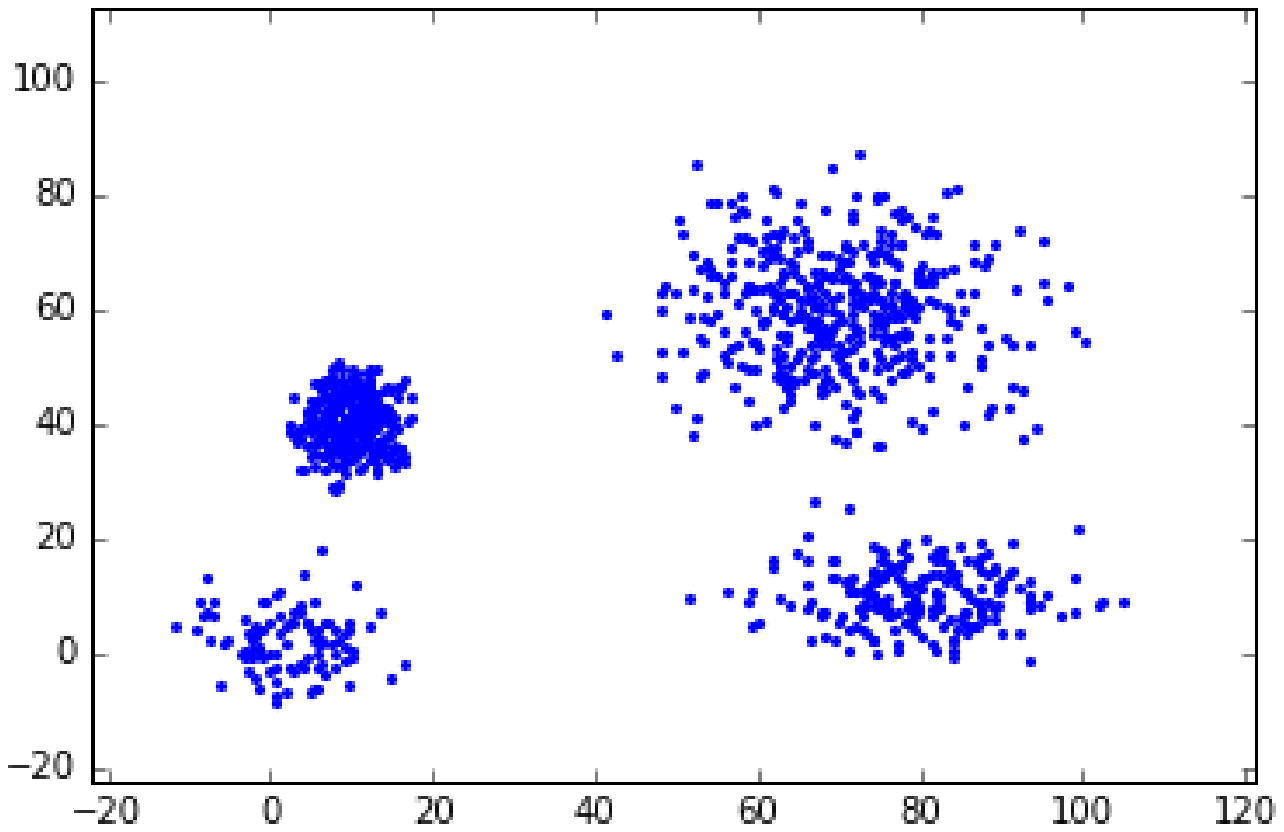}
        \label{fig:CWGAN_GAUSS_RAND_SAMPLE}
	}
    \subfigure[Generated Sample]{
        \includegraphics[width=0.4\textwidth]{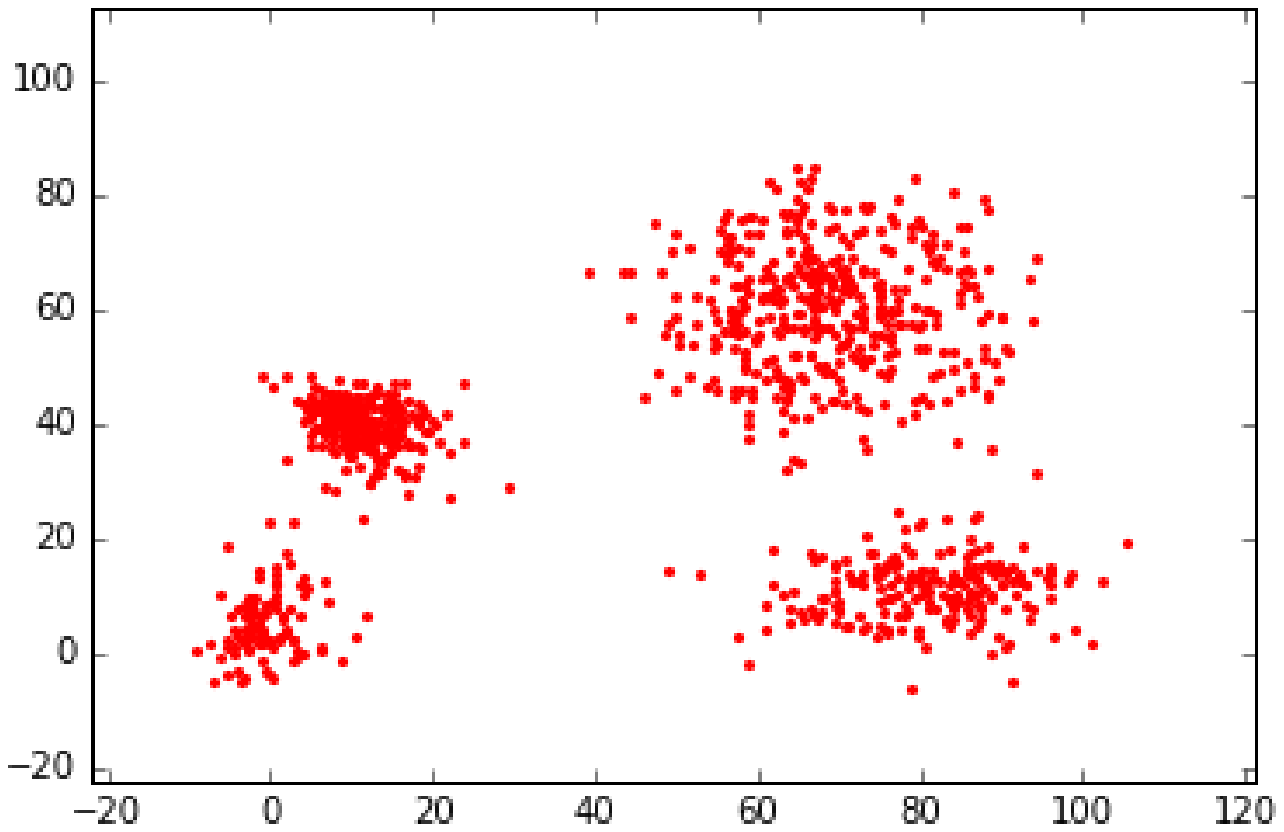}
        \label{fig:CWGAN_GAUSS_GEN_SAMPLE}
	}
	\caption{Random Sample \&.Generated Sample (GAUSS dataset)}
	\label{fig:GAUSSRandomSample&GeneratedSample}
%\vspace{-0.3cm}
\end{figure*}

Fig~\ref{fig:RandomSample&GeneratedSample} and Fig~\ref{fig:GAUSSRandomSample&GeneratedSample} show the distribution of the random samples and generated samples in the first two attributes of the ROAD and GAUSS datasets. The distribution of generated samples and real samples are visually similar in the first two attributes as shown in these figures.
%图中可见，通过GenSample算法能够生成与真实样本数据范围及分布相似的样本。

\subsection{Accuracy}

In this experiment, we compare the accuracies of approximate query results based on the random samples and generated samples.
This experiment is conduced on the ROAD dataset. We test the accuracy of the estimations based on 1,000 random samples and 1,000 generated samples.

We partition the dataset into 10 equi-depth buckets according to the first attribute, suggesting that, the number of tuples in each bucket is the same. We generate 100 samples for each bucket. Thus, we have 1,000 samples in total for the 10 buckets. The number of samples for each group inside a bucket is allocated according to the CVOPT sampling. That is, we adopt Algorithm~\ref{algorithm:CWGAN_CVOPTSample} to generate stratified samples for each bucket.

The query for the $i$th bucket is as follows:

\texttt{SELECT AVG(Elevation), GroupID FROM $Bucket_i$ GROUP BY GroupID;}

\begin{figure*}
	\centering
	\subfigure[Random Sample]{
        \includegraphics[width=1\textwidth]{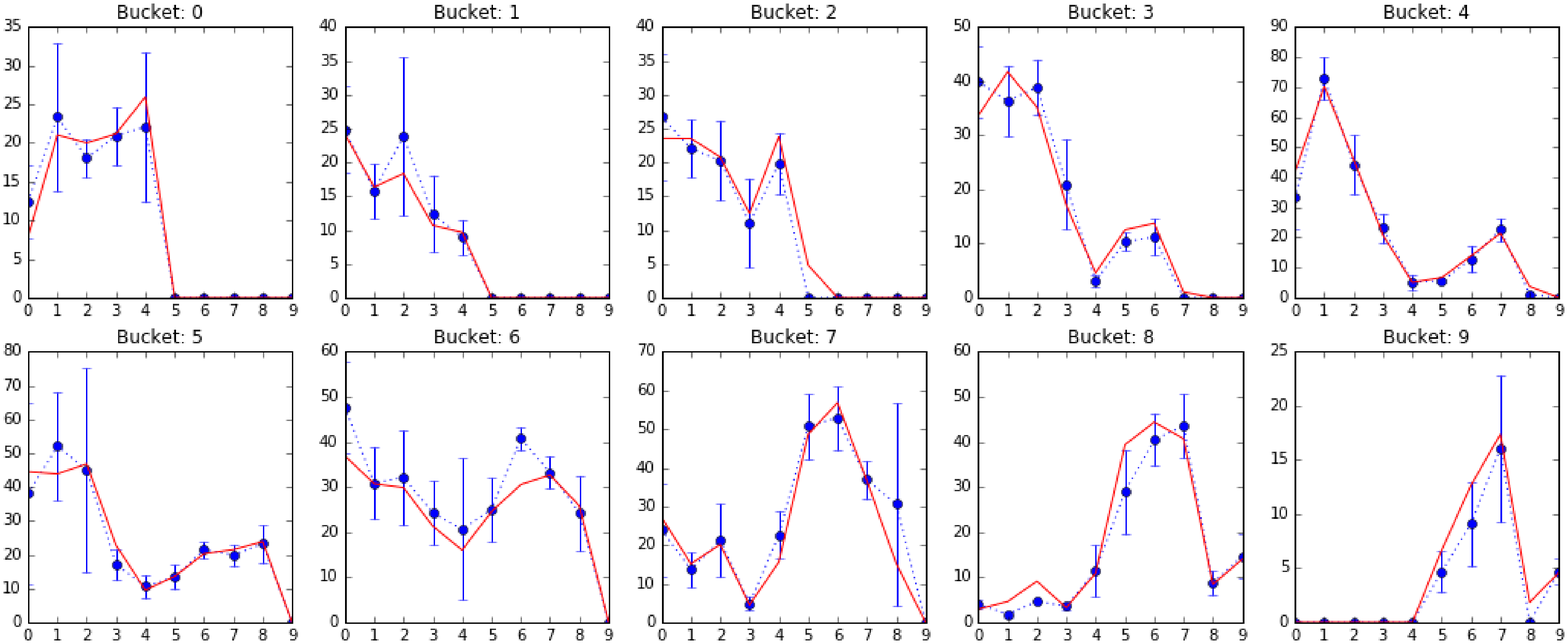}
        \label{fig:CWGAN_RAND_Bucket_DIS}
	}
\centering
    \subfigure[Generated Sample]{
        \includegraphics[width=1\textwidth]{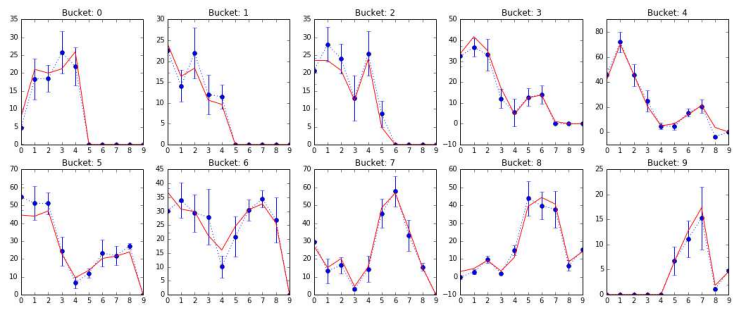}
        \label{fig:CWGAN_GEN_Bucket_DIS}
	}
	\caption{Estimations of group-by queries based on different samples}
	\label{fig:Bucket-Group-By}
%\vspace{-0.3cm}
\end{figure*}

Fig~\ref{fig:Bucket-Group-By} shows the estimations and 95\% confidence intervals based on the random samples and the generated samples. The red line shows the true aggregation results for the 10 groups numbered by 0,1,...,9, and the estimations are shown as the blue dotted lines with error bars. We can learn from the figure that the confidence intervals of the estimations based on the random samples vary greatly for different groups due to the skew of the data. However, the accuracies of the estimations based on the generated samples seem to be more stable among different groups. The reason is that we combine the generative model with the CVOPT sampling to generate stratified samples for each group. The experimental results show that, the conditional generative model can be easily combined with the stratified sampling method to increase the accuracy. In addition, the generating samples based on the model have the advantage over the stratified sampling, since stratified sampling requires accessing the entire dataset, while the generated sample does not.

\subsection{Generative-model-based Online-Aggregation}
In this experiment, we show that the generative model can be applied to the online-aggregation, which continuously narrows down the confidence interval and increases the accuracy by accessing more samples.

\begin{figure*}
	\centering
	\subfigure[Impact of sample size on confidence Interval]{
        \includegraphics[width=0.4\textwidth]{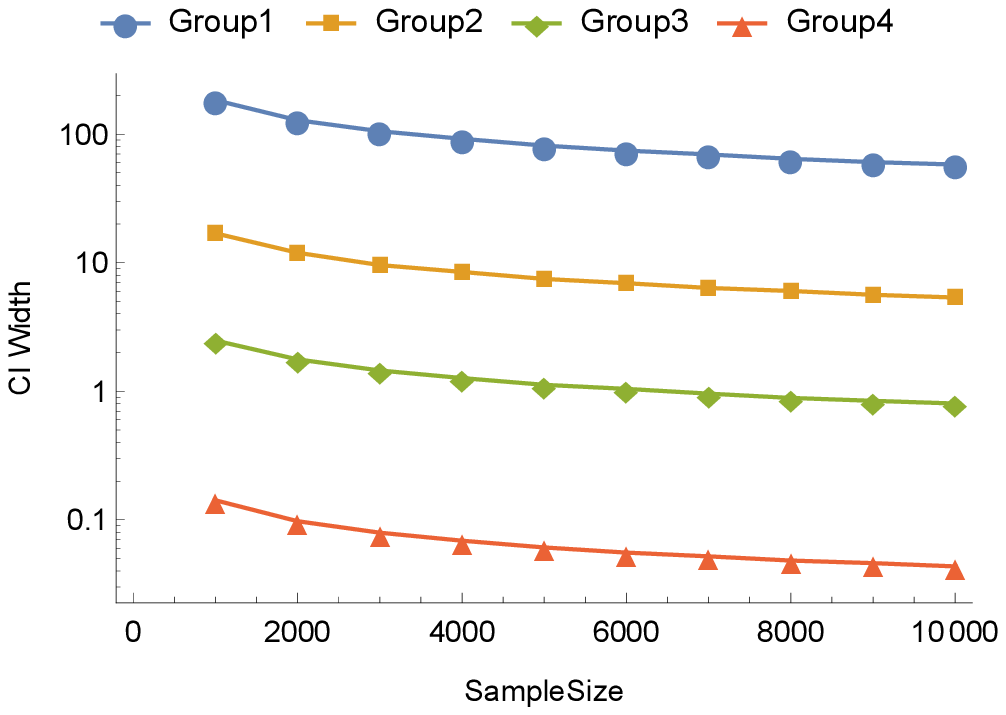}
        \label{fig:CWGAN_CIWidth}
	}
    \subfigure[Impact of sample size on $\frac{CI}{Mean}$]{
        \includegraphics[width=0.4\textwidth]{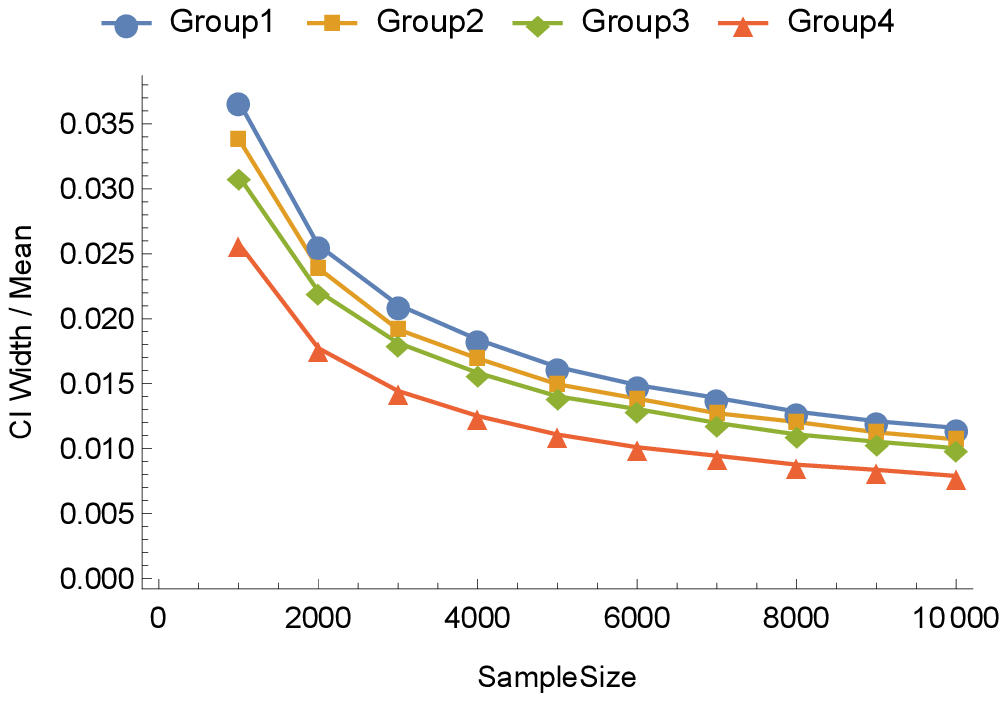}
        \label{fig:CWGAN_CI_Mean}
	}
	\caption{Online Aggregation based on Generative model}
	\label{fig:CWGAN_OnlineAggregation}
%\vspace{-0.3cm}
\end{figure*}

This experiments are conducted on the GAUSS dataset. For each group, we set the sample sizes from 1,000 to 10,000 to test the accuracy of the samples in different sizes. Fig~\ref{fig:CWGAN_OnlineAggregation} shows the confidence intervals of the estimations based on the samples in different sizes. It is clear that the confidence interval narrows down as the sample size increases. Fig~\ref{fig:CWGAN_CI_Mean} measures the accuracy according to $\frac{ConfidenceInterval}{Mean}$, which limits the relative error of the estimations. The accuracy also increases with the sample size. The results indicate that the generated samples can also be adopted to online-aggregation and it can narrow down the confidence interval and increase the accuracy by generating more samples.

\subsection{The impact of selectivity on the accuracy}
In this experiment we compare the accuracy of different methods and test the impact of selectivity on the accuracy.

We compare the accuracy of the estimations based on the generated samples, the random samples, and the DBEst. DBEst is the state-of-the-art ML-model-based AQP method, which builds the probability density models according to the samples from data.

This experiment is conducted on the ROAD dataset.
We test the accuracies of 100 queries whose selectivities are below 0.05. The queries are in the following form:

\texttt{SELECT AVG(Elevation), GroupID FROM ROAD,}

\texttt{WHERE a $\le$ Longitude $\le$ b GROUP BY GroupID;}

\begin{figure*}
	\centering
	\subfigure[MSE]{
        \includegraphics[width=0.4\textwidth]{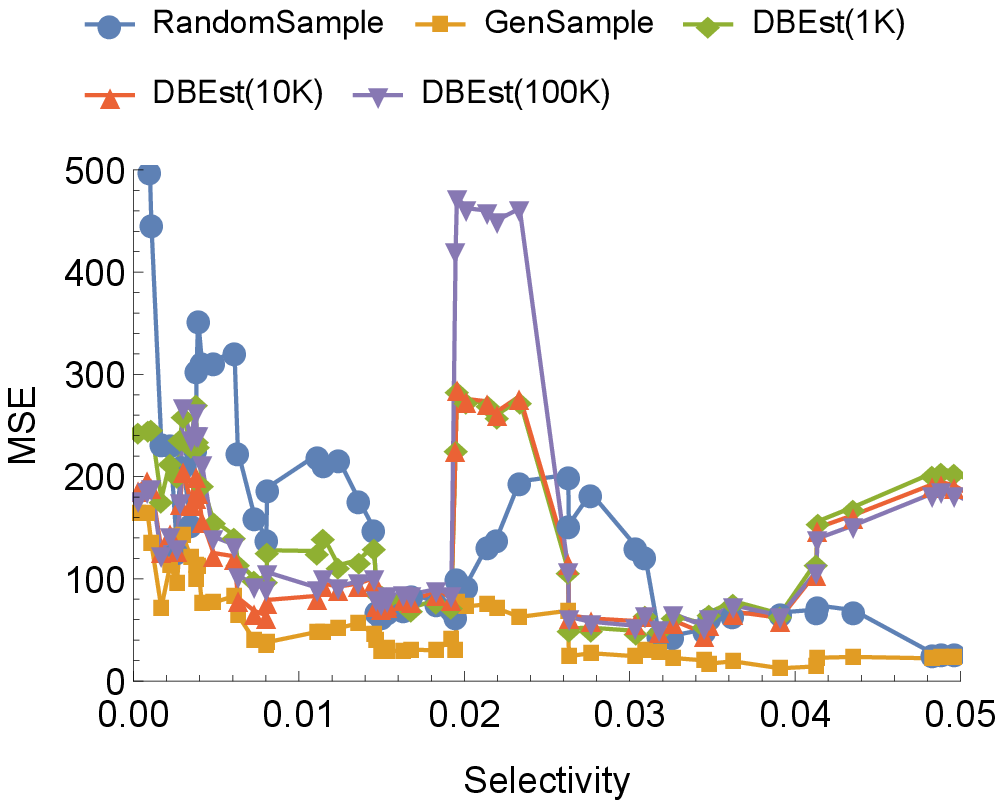}
        \label{fig:CWGAN_Selectivity_MSE}
	}
    \subfigure[ARE]{
        \includegraphics[width=0.4\textwidth]{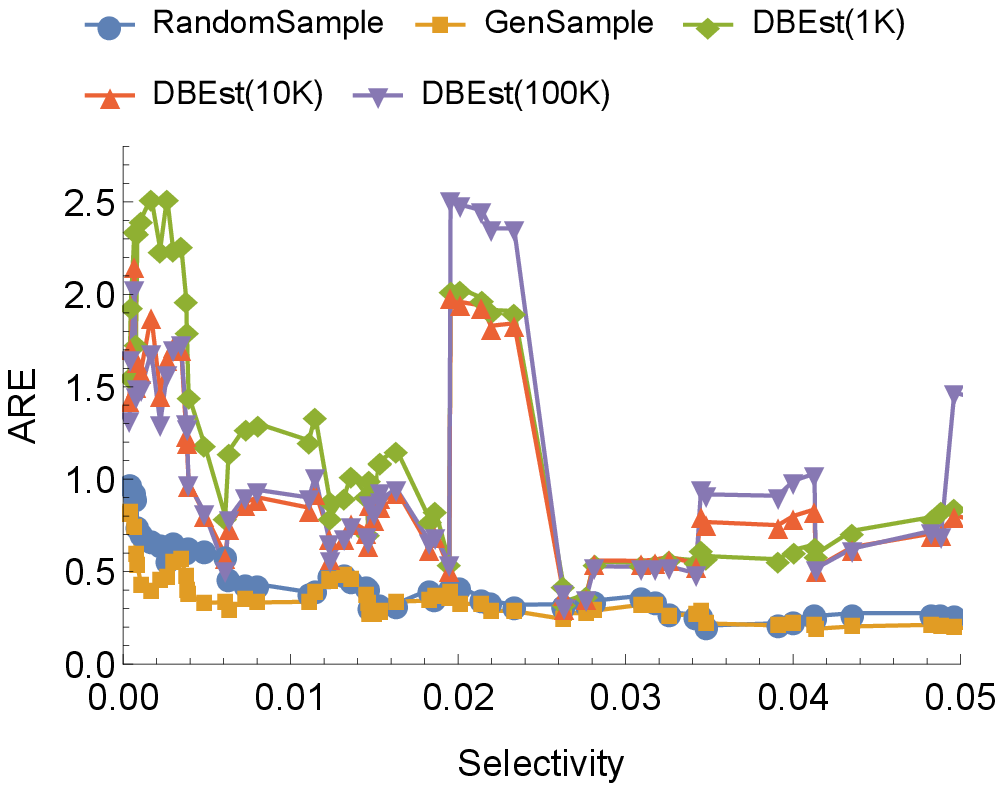}
        \label{fig:CWGAN_Selectivity_ARE}
	}
	\caption{The impact of selectivity on the accuracy}
	\label{Fig:CWGAN_Selectivity}
%\vspace{-0.3cm}
\end{figure*}

We choose 1,000 random samples from the dataset. We adopt Algorithm~\ref{algorithm:CWGAN_SampleforPredicates} to generate 1,000 samples for each query according to its predicate. We randomly choose 1k, 10k, 100k samples to build DBEst models, respectively.
The result is shown in Fig~\ref{Fig:CWGAN_Selectivity}. For both the random samples and generated samples, the estimation error decreases as the query selectivity increases, and the estimations based on the generated samples are more accurate. The reason is that the generated samples contains a higher proportion of the samples satisfying the query predicates, compared with the random samples chosen from the entire dataset. The queries in this experiment are low-selectivity queries, meaning that, most of the samples chosen from the entire dataset do not satisfy the query predicates. Only a small part of the samples contribute to the estimation, which leads to low accuracy.
Since the generative model can flexibly generate samples for some sub-ranges, it can obtain more samples that contribute to the estimation. Since more samples leads to more accurate estimations, the generated samples can bring higher accuracy for a low-selectivity query.
The accuracy of DBEST has not flattened out with the increased selectivity, which may be caused by the inaccurate probability density model built according to the samples.

\subsection{The impact of data size on the sampling efficiency}

\begin{figure}
\centering
\includegraphics[width = 0.4\textwidth]{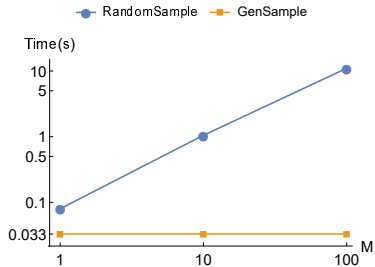}
\caption{The impact of data size on the sampling time}
\label{Fig:CWGAN_efficiency}
\end{figure}

In this algorithm, we compare the efficiency of generating samples and choosing random samples from the dataset.
This experiment is conducted on the GAUSS dataset. We compare the sampling time of 1,000 samples from 1M, 10M, and 100M datasets with the time of generating 1,000 samples from the model. The results are shown in Fig~\ref{Fig:CWGAN_efficiency}.
The sample generation time is undoubtedly independent of the data size, but the random sampling time is linearly related to the data size. That is, generating samples from the model is more efficient than sampling from a large amount of data.

\subsection{Space Cost}

\begin{figure}
\centering
\includegraphics[width = 0.4\textwidth]{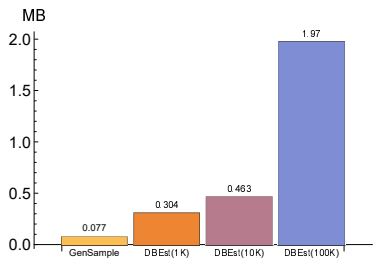}
\caption{The Space Cost of the models in GenSample and DBEst}
\label{Fig:CWGAN_Space}
\end{figure}

In this experiment, we compare the space cost of the generative model with that of the DBEst. We test the space cost of the models in DBEst built from 1k, 10k, and 100k samples. The experimental result is shown in Fig~\ref{Fig:CWGAN_Space}. The space cost of the generative model built according to the implementation introduced in Section~\ref{section:CWGAN-implement} is 77KB. The figure shows that the space cost of the generative model is less than the DBEst. The reason is that, DBEst separately builds a model for each group, while different groups share the same model in our method. Thus, the conditional generative model in our framework is more compact than the DBEst.

\section{Conclusions}\label{section:conclusion}
In this paper, we proposed a generative-model-based approximate query processing framework for group-by queries. Samples in each group can be generated from the conditional generative model without accessing the raw data, which avoids the latency caused by sampling from the large amount of data. The proposed framework can collaborate with the uniform sampling to restore the data distribution, and collaborate with the stratified sampling and online aggregation to improve the accuracy.
In this work, we separate the data into buckets to build the conditional generative model which can generate targeted samples for each small bucket. We will try to make the conditional generative model directly generate samples for a given range in our future study.

\bibliography{mybibfile}

\end{document}